\documentclass[aps,prl,reprint]{revtex4-2}
\usepackage{blindtext}
\usepackage{graphicx}
\usepackage{subcaption}
\usepackage{xcolor}
\usepackage[fleqn]{amsmath}
\usepackage[font=small,skip=3pt]{caption}
\usepackage{makecell}
\usepackage{hyperref} 
\usepackage{cleveref}
\usepackage{url}
\begin{document}
\title{Kinetic-Ballooning-Bifurcation in Tokamak Pedestals Across Shaping and Aspect-Ratio}
\author{J. F. Parisi$^{1}$}
\email{jparisi@pppl.gov}
\author{A. O. Nelson$^2$} 
\author{R. Gaur$^3$}
\author{S. M. Kaye$^1$}
\author{F. I. Parra$^1$}
\author{J. W. Berkery$^1$}
\author{K. Barada$^4$}
\author{C. Clauser$^5$}
\author{A. J. Creely$^6$}
\author{A. Diallo$^1$}
\author{W. Guttenfelder$^{1,7}$}
\author{J. W. Hughes$^5$}
\author{L. A. Kogan$^8$}
\author{A. Kleiner$^1$} 
\author{A. Q. Kuang$^6$} 
\author{M. Lampert$^1$}
\author{T. Macwan$^4$}
\author{J. E. Menard$^1$}
\author{M. A. Miller$^5$}
\affiliation{$^1$Princeton Plasma Physics Laboratory, Princeton University, Princeton, NJ, 08540, USA}
\affiliation{$^2$Department of Applied Physics and Applied Mathematics, Columbia University, New York, NY,  10027, USA}
\affiliation{$^3$Department of Mechanical and Aerospace Engineering, Princeton University, Princeton, NJ, 08540, USA}
\affiliation{$^4$University of California, Los Angeles, Los Angeles, CA, 90095, USA}
\affiliation{$^5$Plasma Science and Fusion Center, Massachusetts Institute of Technology, Cambridge, MA, 02139, USA}
\affiliation{$^6$Commonwealth Fusion Systems, Devens, MA, 01434, USA}
\affiliation{$^7$Type One Energy, 8383 Greenway Boulevard, Middleton, WI, 53562, USA}
\affiliation{$^8$United Kingdom Atomic Energy Authority, Culham Science Centre, Abingdon, OX14 3DB, UK}

\begin{abstract}
We {use} a new gyrokinetic threshold model to predict a bifurcation in tokamak pedestal width-height scalings that depends strongly on plasma shaping and aspect-ratio. {The} bifurcation arises from the first and second stability properties of kinetic-ballooning-modes that {yields} wide and narrow pedestal branches, {expanding} the space of accessible pedestal widths and heights. The wide branch offers potential for edge-localized-mode-free pedestals with high core pressure. For negative triangularity, low-aspect-ratio configurations are predicted to give steeper pedestals than conventional-aspect-ratio. Both wide and narrow branches have been attained in tokamak experiments.
\end{abstract}

\maketitle

\setlength{\parskip}{0mm}
\setlength{\textfloatsep}{5pt}

\setlength{\belowdisplayskip}{6pt} \setlength{\belowdisplayshortskip}{6pt}
\setlength{\abovedisplayskip}{6pt} \setlength{\abovedisplayshortskip}{6pt}

The realization of magnetic confinement fusion energy represents a significant milestone in the quest for clean and abundant power sources. This endeavor hinges on the ability to confine a high-pressure plasma {with} magnetic field{s}, as characterized by the ratio $\beta = 2 \mu_0 p/ B^2$, where $p$ is the plasma pressure and $B$ the magnetic field strength. For tokamaks operating in high-confinement mode (H-mode), an intriguing phenomenon {occurs}—the formation of a pedestal at the plasma edge \cite{Wagner1982}, often enhancing $\beta$ substantially and hence fusion power. However, ballooning modes pose a fundamental challenge to achieving optimal $\beta$ values. Recent experiments in negative triangularity plasmas \cite{Austin2019, Nelson2023} showed that plasma shaping changes pedestal ballooning stability substantially. In this Letter, we describe a bifurcation in the H-mode pedestal width and height that can be manipulated and optimized with plasma shaping and aspect-ratio. The bifurcation arises from ballooning stability {properties} and {presents} new pedestal operating scenarios for reactors. By degrading ballooning stability in the edge, higher core $\beta$ may be achieved by enabling the pedestal to grow without triggering edge-localized-modes (ELMs) that threaten plasma-facing components \cite{Osborne2015,Snyder2009}. 

Predictive models {constrain} the pedestal {radial width} $\Delta_{\mathrm{ped}}$ and height $\beta_{\theta,\mathrm{ped}}$ by ballooning and ELM stability \cite{Snyder2009, Snyder2011}.  Microscopic ballooning stability gives a width-height scaling that constrains pedestal pressure gradients. The macroscopic ELM constraint gives a `hard' limit that transports significant {pressure} once triggered \cite{Snyder2002, Wilson2004}. These two constraints intersect to give a $\Delta_{\mathrm{ped}}$, $\beta_{\theta,\mathrm{ped}}$ prediction. Previous work has shown that the ELM constraint is sensitive to plasma shaping: negative triangularity TCV tokamak plasmas were subject to a much more restrictive ELM constraint than positive triangularity \cite{Merle2017}, whereas on the DIII-D tokamak, fueling and strong positive triangularity provided a route to `Super H-mode' \cite{Snyder2015,Snyder2019} pedestals. The new bifurcation in this work {complements these results}, focusing on kinetic-ballooning-mode (KBM) \cite{Tang1980} physics.

{To find pedestal width-height scalings, we parameterize radial electron temperature and density pedestal profiles with $\tanh$ functions \cite{Mahdavi2003,Snyder2009} so that pedestal structure can be varied meaningfully with two variables,}
{
\begin{equation}
\Delta_{\mathrm{ped}} = (\Delta_{n_e} + \Delta_{T_e})/2, \;\;\; \beta_{\theta, \mathrm{ped}} = 2 \mu_0 p_{\mathrm{ped}} /\overline{B}_{\mathrm{pol}}^2.
\end{equation}
}{Here, $\Delta_{n_e}$ and $\Delta_{T_e}$ are the electron density and temperature pedestal widths in units of normalized poloidal flux, $p_{\mathrm{ped}} = 2 p_e (\psi = \psi_{\mathrm{ped}} )$ where $\psi_{\mathrm{ped}}$ is the poloidal flux $\psi$ at the pedestal top, and $\overline{B}_{\mathrm{pol}} = \mu_0 I_p /  l $ with last-closed-flux-surface circumference $l$ \cite{Snyder2009,Smith2022}. The width-height scaling is the boundary in $\Delta_{\mathrm{ped}}$, $\beta_{\theta,\mathrm{ped}}$ space separating pedestals that are limited by KBMs. For more details, see \cite{Snyder2009,Parisi2024stability}. Experimentally, $\Delta_{\mathrm{ped}}$ and $ \beta_{\theta, \mathrm{ped}}$ typically grow close to the width-height scaling trajectory \cite{Snyder2012}, facilitated by a radially broadening flow shear profile that suppresses long-wavelength turbulence. For ELMy H-modes, growth ceases when peeling-ballooning-modes (PBMs) are destabilized, causing an ELM \cite{Connor1998,Wilson1999,Kirk2004,Snyder2009}.}

\begin{figure}[!tb]
    \centering
    \includegraphics[width=0.5\textwidth]{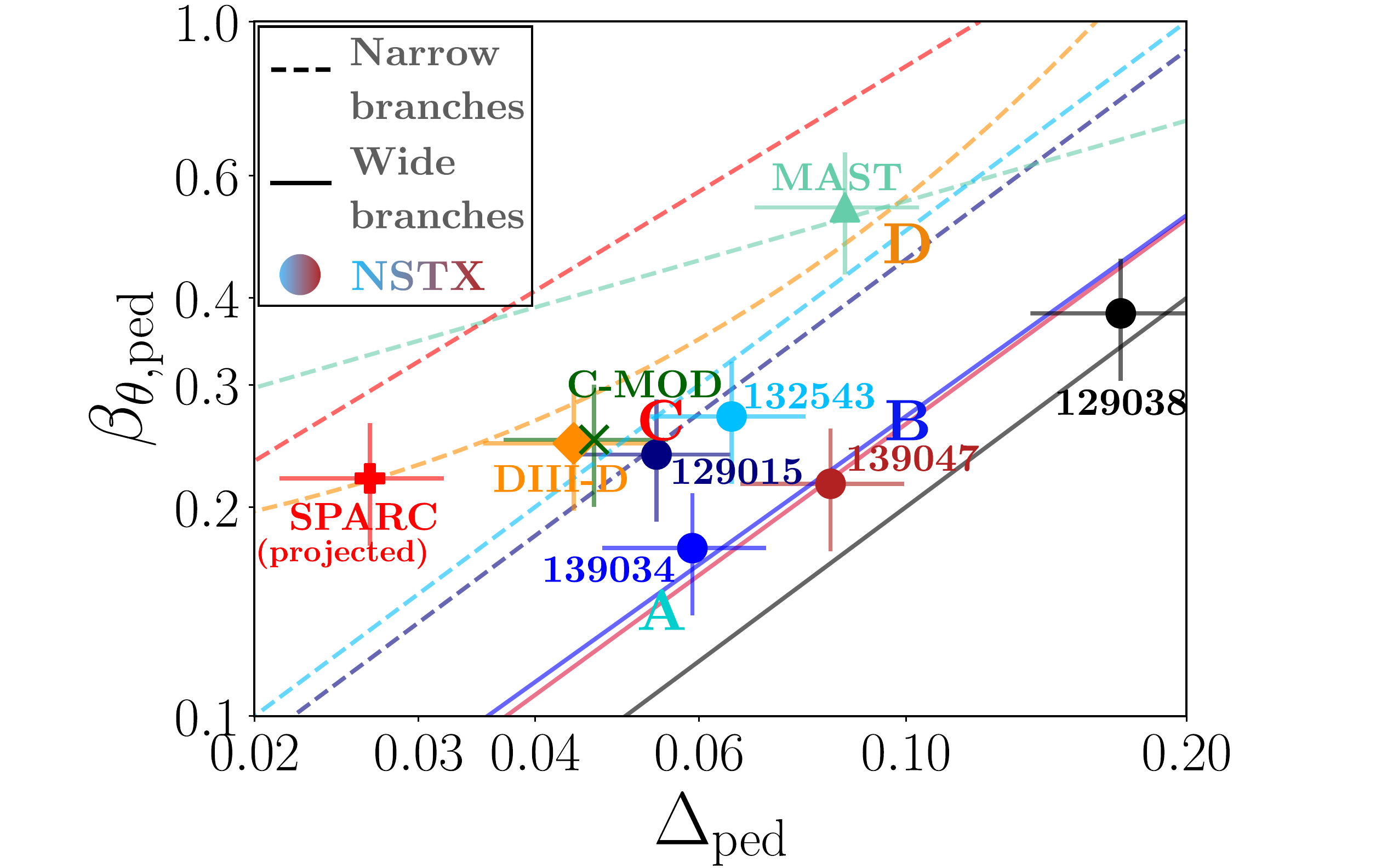}
    \caption{{Pedestal width-height bifurcation} for DIII-D, MAST, NSTX, and SPARC discharges, with width-height scalings of the branch (wide or narrow) closest to the experimental point. NSTX discharge numbers are indicated next to circle markers, DIII-D \#163303, MAST \#29782, C-Mod \#1101214029. {A-D are points on NSTX 139047 wide and narrow (not shown) branches (see \cref{fig:one} for $s-\alpha$ stability). Errorbars are 20\% of $\Delta_{\mathrm{ped}}$, $\beta_{\theta, \mathrm{ped}}$ values.}}
    \label{fig:zero}
\end{figure}

The KBM has two stability branches \cite{Greene1981}. {At relatively low pressure gradients, the normalized gradient}
\begin{equation}
{\alpha = (\mu_0/2\pi^2)(\partial V / \partial \psi) \sqrt{V/(2\pi^2 R_0)} (dp/d\psi)},
\end{equation}
increases until destabilizing pressure gradients overcome the stabilizing effects of field-line bending \cite{Connor1979}. Here, $V$ is the flux-surface enclosed volume and $R_0$ the plasma major radius. At much higher $\alpha$, the local magnetic shear at the low-field side becomes highly negative and stabilizing, which leads to second stability \cite{Greene1981}. A key result of this work is that first and second KBM stability across the pedestal is realized as two distinct branches -- radially wide and narrow -- of ballooning width-height scalings. {For ELMy H-modes, given that the ELM limit is related to stored pedestal energy \cite{Loarte2003}, lower gradient pedestals are intrinsically `wider' and steeper gradient pedestals are `narrower.' We will also show that wide pedestals offer  potential for robust ELM-free operation.}

In \cref{fig:zero}, we show the pedestal width-height bifurcation for National-Spherical-Torus-Experiment (NSTX) discharges 129015, 129038, 132543, 139034, and 139047 \cite{Maingi2012,Diallo2013}. {The width-height scaling lines in \cref{fig:zero} are obtained with a linear KBM threshold model \cite{Parisi2023_ARXIV} using GS2 gyrokinetic simulations \cite{Dorland2000}.} Each NSTX discharge admitted two solutions: a wide and narrow branch. However, 139034, 139047, and 129038 equilibria -- indicated by markers in \cref{fig:zero} -- lie on the wide branch, whereas 129015 and 132543 lie on the narrow branch. These discharges demonstrate that NSTX experiments {occupy} both wide and narrow pedestal branches. For readability, in \cref{fig:zero} we only plot {branches} closest to the {experimental point}. We also plot equilibrium points and width-height scalings for DIII-D 163303 \cite{Grierson2018}, {lying on} the narrow branch, MAST 29782 \cite{Smith2022}, also on the narrow branch, and {a case representing the Primary Reference Discharge for} SPARC \cite{Creely2020, Hughes2020, CFS2022}, whose operational point is projected below the narrow branch. While we do not calculate a scaling, we show the {ELMy H-mode pedestal} C-Mod 1101214029 equilibrium point \cite{Hughes2013, Groebner2013} lies in the narrow branch region.

\begin{figure}[!tb]
    \centering
    \includegraphics[width=0.17\textwidth]{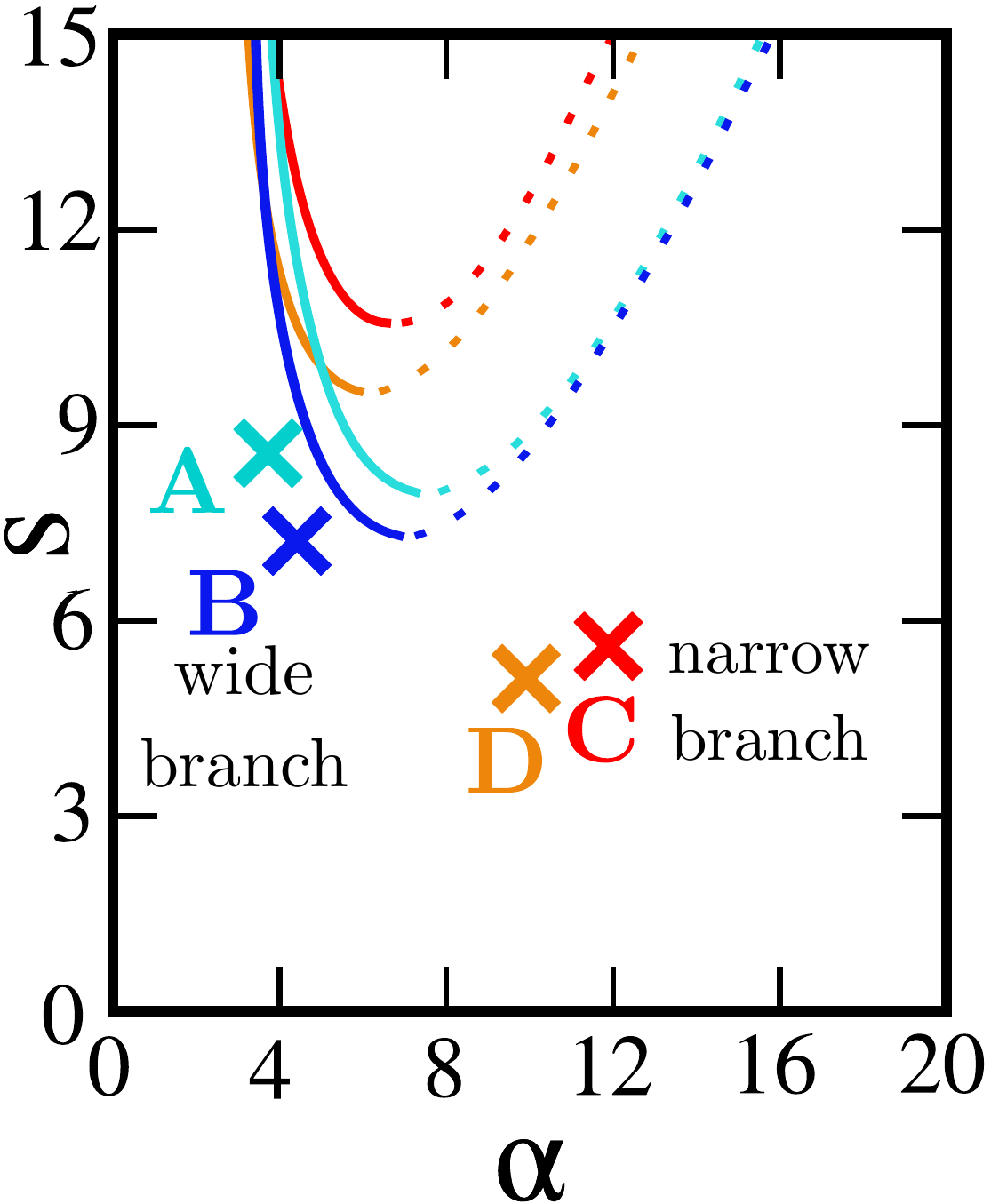}
    \caption{{Ideal $s-\alpha$ stability for points A-D in \cref{fig:zero}. Points A and B (C and D) are in the wide (narrow) ballooning branch and are evaluated at the pedestal mid-radius.}}
    \label{fig:one}
\end{figure}

{We perform ideal $s-\alpha$ analysis \cite{Greene1981} in \cref{fig:one} at the pedestal mid-radius for equilibria on the wide and narrow branches of NSTX 139047, shown by A, B and C, D in \cref{fig:zero} (we omit the 139047 narrow branch scaling in \cref{fig:zero}). We plot ideal stability curves versus $\alpha$ and magnetic shear $s = (r/q)(dq/dr)$ where $r$ is the flux-surface half-diameter. The wide branch points A, B are in the ideal first-stable region, and the narrow branch points C, D are ideal second-stable below the stability curve `nose.' Because the scalings in \cref{fig:zero} use gyrokinetic stability, points A-D are relatively far from the ideal boundaries.}

\begin{figure}[!b]
    \centering
    \begin{subfigure}[t]{0.15\textwidth}
    \centering
    \includegraphics[width=1.3\textwidth]{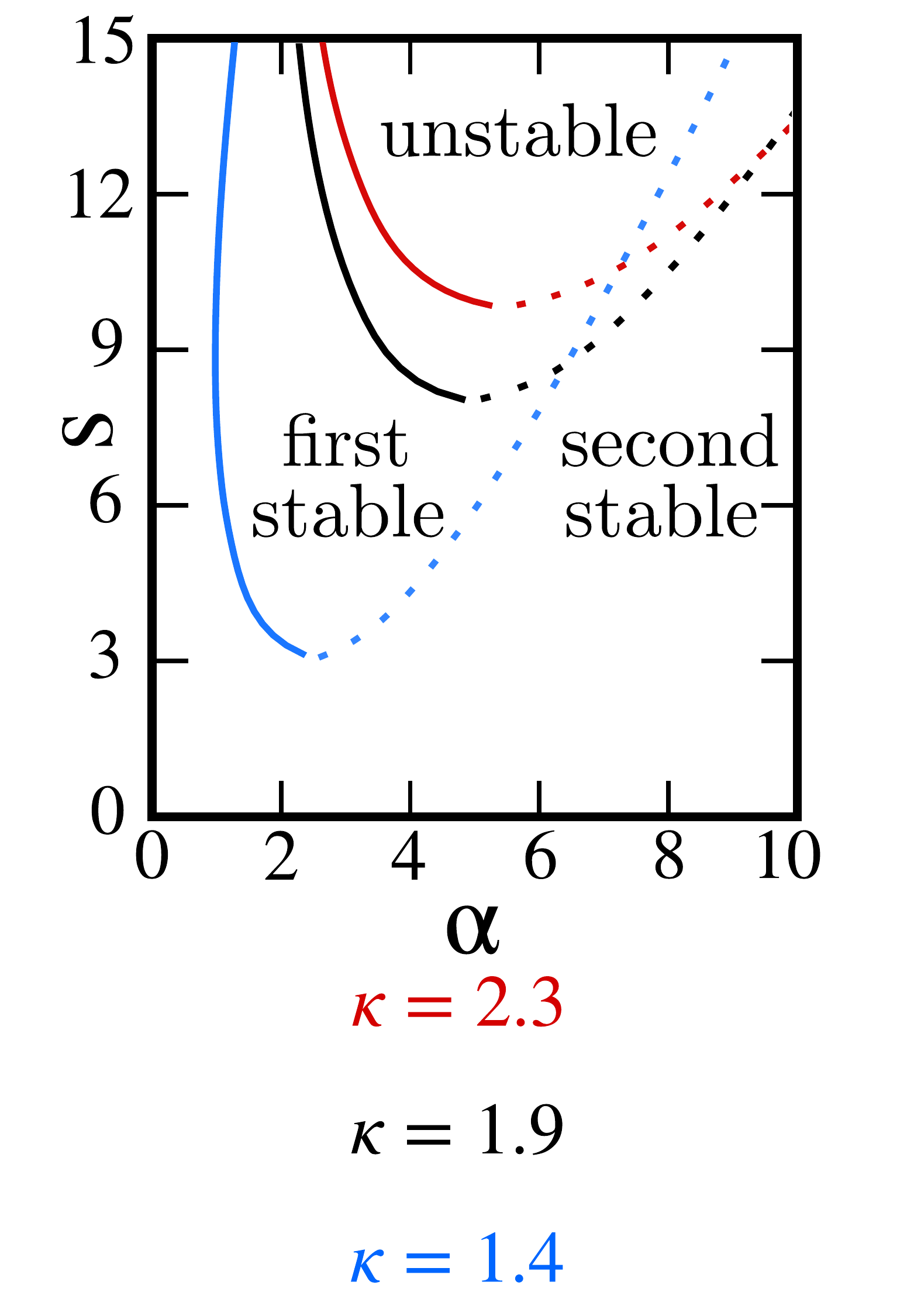}
    \caption{\textcolor{black}{Elongation}}
    \end{subfigure}
     ~
    \begin{subfigure}[t]{0.15\textwidth}
    \centering
    \includegraphics[width=1.3\textwidth]{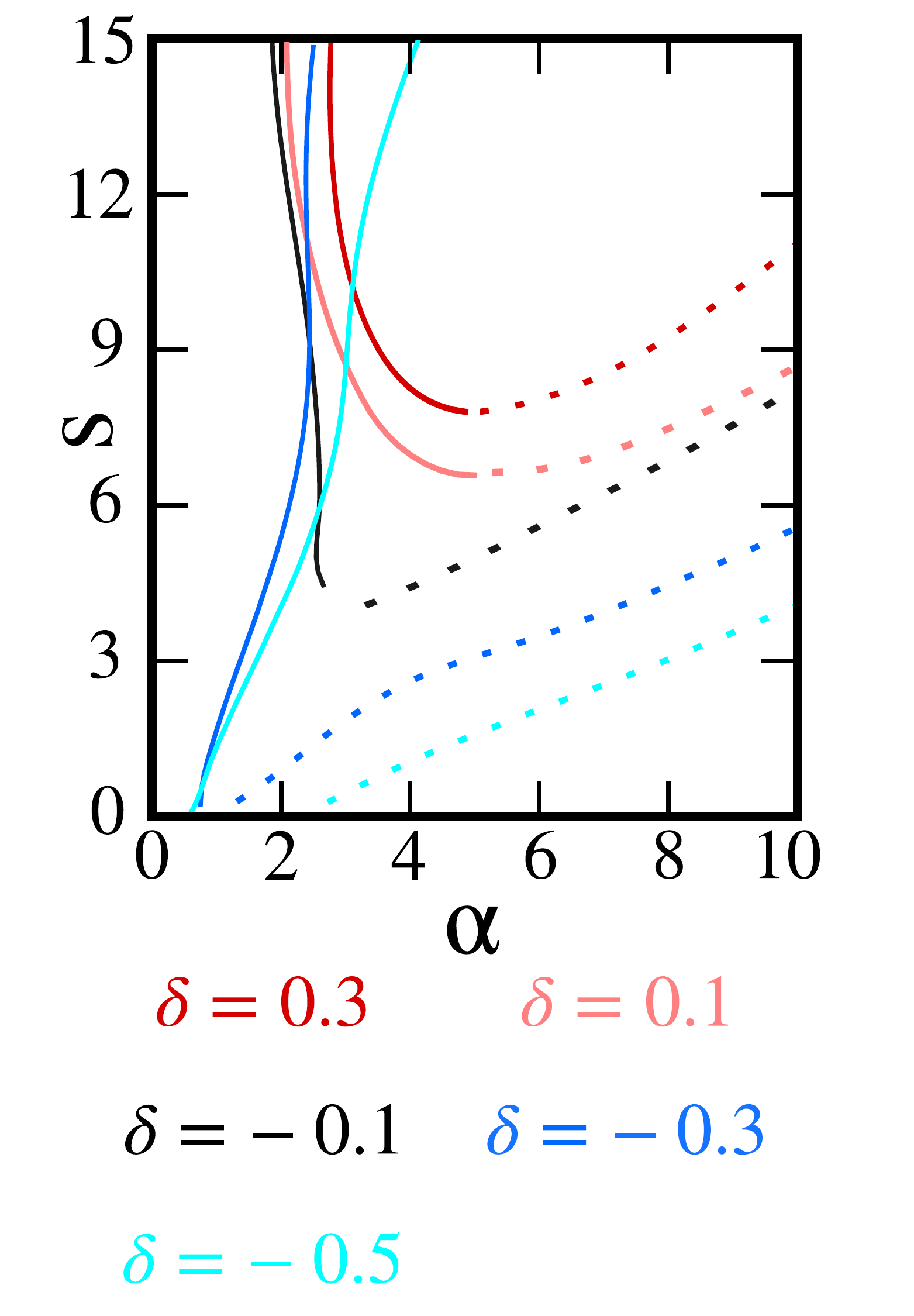}
    \caption{\textcolor{black}{Triangularity}}
    \end{subfigure}
     ~
    \begin{subfigure}[t]{0.15\textwidth}
    \centering
    \includegraphics[width=1.3\textwidth]{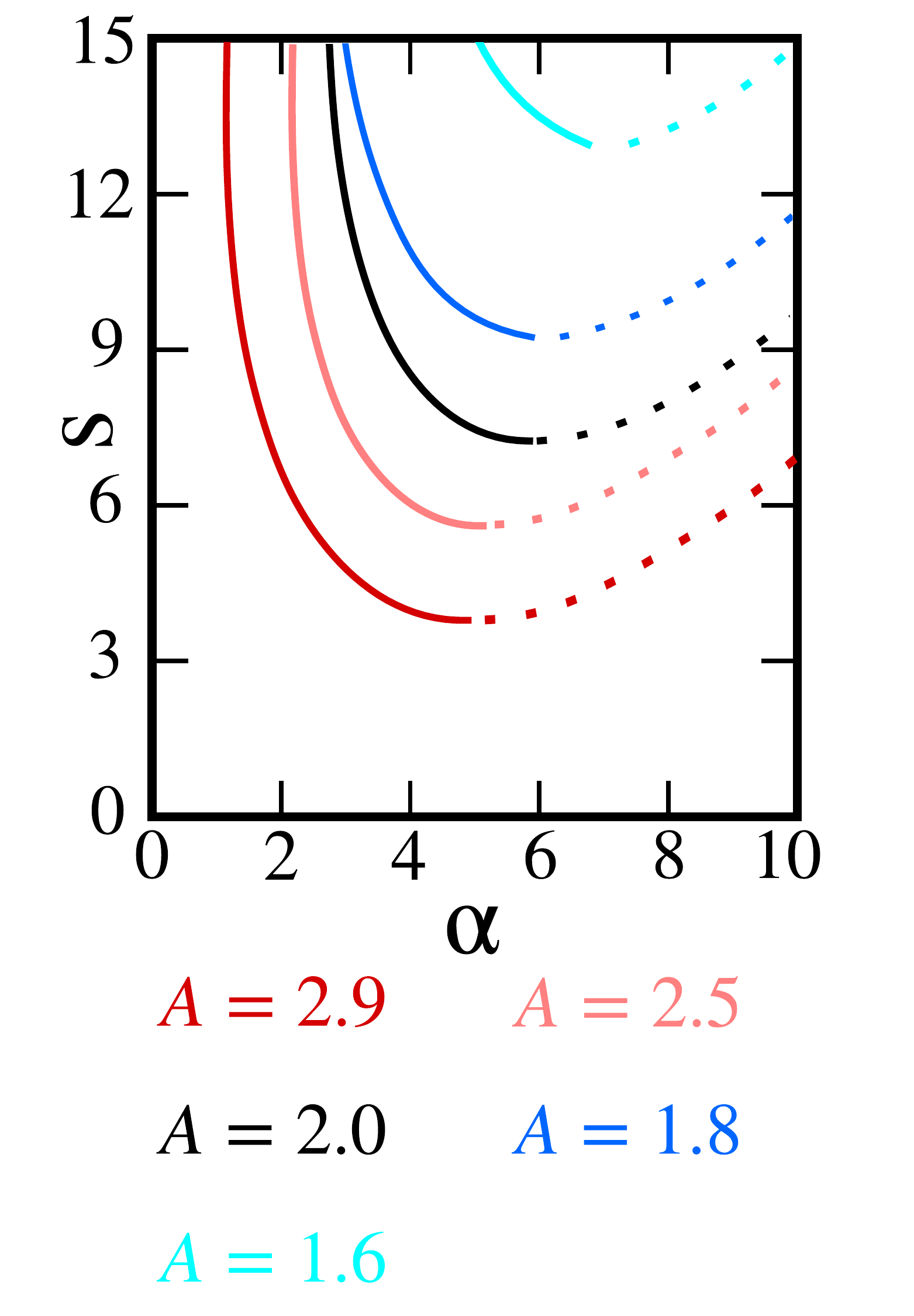}
    \caption{\textcolor{black}{Aspect-Ratio}}
    \end{subfigure}
    \caption{Ideal $s-\alpha$ stability for different $\kappa$, $\delta$, and $A$ values for the nominal pedestal width and height at the {pedestal mid-radius} for NSTX 139047. First and second stability curves are solid and dashed, respectively. In (a), first-stable, second-stable, and unstable regions are indicated for $\kappa = 1.9$.}
    \label{fig:salpha_tri_elong_aspect}
\end{figure}

\begin{figure*}[!tb]
    \centering
    \includegraphics[width=0.8\textwidth]{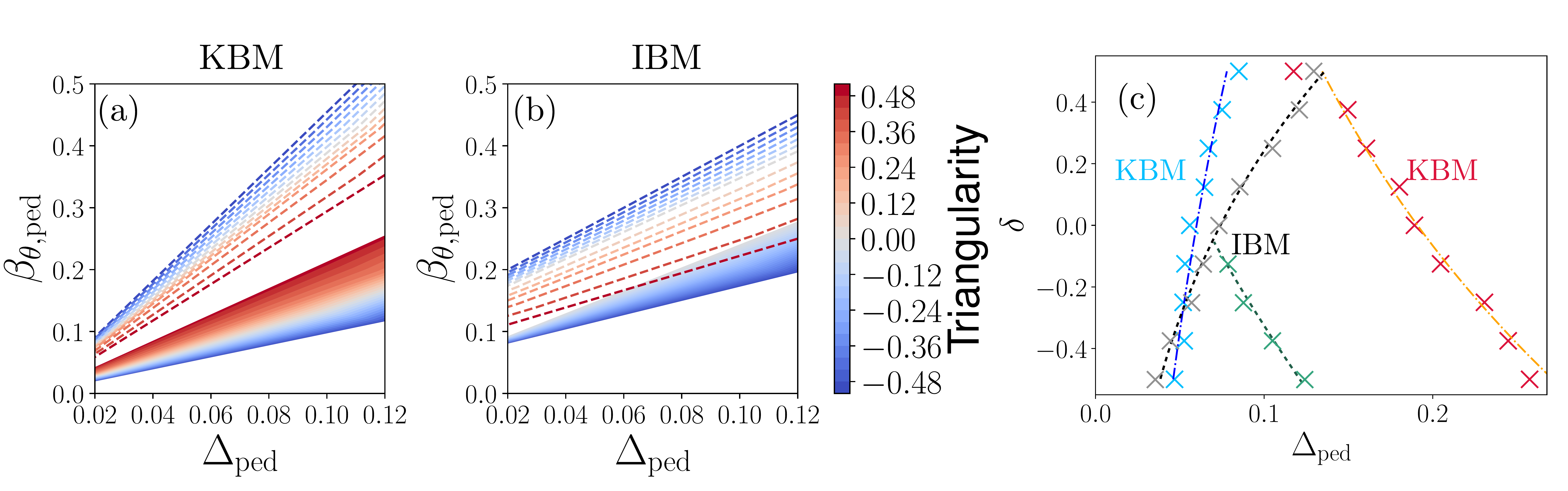}
    \caption{Width-height scalings across triangularity using (a) the kinetic-ballooning-mode (KBM) and (b) the ideal-ballooning-mode (IBM). In (c), we take a constant $\beta_{\theta, \mathrm{ped}} = 0.25$ slice and plot the {fits (see \Cref{eq:deltascalings_tri}) for} triangularity $\delta$ versus pedestal width $\Delta_{\mathrm{ped}}$. A bifurcation exists for all values of $\delta$ in the KBM scalings, but not for the IBM scalings.}
    \label{fig:four}
\end{figure*}

We now demonstrate how plasma shaping and aspect-ratio enter the ballooning bifurcation. The geometric shape of the last-closed-flux-surface and {shaping's} penetration to the magnetic axis {impacts} plasma performance \cite{Goldston1984, Kaye1985, Ball2015, Austin2019} and {changes} ballooning mode stability \cite{Turnbull1999, Nelson2022}. Starting from an experimental equilibrium for NSTX 139047, we vary the shaping parameters and aspect-ratio. We use {Luce parameters} for the plasma shape \cite{Luce2013}, defining the elongation $\kappa$ and triangularity $\delta$ as the average $\langle \ldots \rangle_L$ of the Luce parameters $\kappa = \langle \kappa \rangle_L, \;\; \delta = \langle \delta \rangle_L$. When varying $\delta$, the total plasma current $I_p$ is held constant.  {Additionally, due to its impact on ballooning stability \cite{Connor2016}, the quantity $\beta_N = 2 \mu_0 \langle p \rangle a / I_p B_{T0}$ is held constant.} Here, $\langle p \rangle$ is the volume-averaged pressure, $a$ the minor radius, and $B_{T0}$ the toroidal magnetic field strength at the magnetic axis. When varying $\kappa$, we scale $I_p \sim 1 + \kappa^2$, and when varying aspect-ratio $A = R_0/a$, we scale $\beta_N \sim 1/\sqrt{R_0}, \;\; B_{T0} \sim R_0, \;\; I_p \sim \sqrt{R_0}$ \cite{Menard2016}.

We {first} describe the effect of shaping and aspect-ratio on ideal-ballooning-mode stability {(note that ideal and kinetic ballooning stability boundaries can differ, often significantly \cite{Tang1980, Hastie1981, Dong1999, Pueschel2008, Ma2017, Aleynikova2017, Parisi2023_ARXIV}).} For NSTX 139047, we compare equilibria with different {$\kappa$, $\delta$, and $A$ values} {with} ideal $s-\alpha$ analysis. In \cref{fig:salpha_tri_elong_aspect} (a), increasing elongation from $\kappa = 1.4$ to $2.3$ shifts the first-stable boundary to much higher $s$ and $\alpha$. This suggests that a KBM first-stable width-height scaling branch will be much wider radially {for a similar pedestal height (i.e. larger $\Delta_{\mathrm{ped}}$)} at lower elongation because the first-stable boundary ({in $s-\alpha$ space}) is at lower $\alpha$ values. Similarly in \cref{fig:salpha_tri_elong_aspect} (b), decreasing triangularity to negative values decreases the $s$ and $\alpha$ values for the first-stable boundary, suggesting that a KBM first-stable, wide branch will be much wider at lower and negative triangularity. Finally, \cref{fig:salpha_tri_elong_aspect} (c) suggests that the first-stable branch will be widest at high-aspect-ratio.

\begin{figure}[!b]
    \centering
    \begin{subfigure}[t]{0.231\textwidth}
    \centering
    \includegraphics[width=1.09\textwidth]{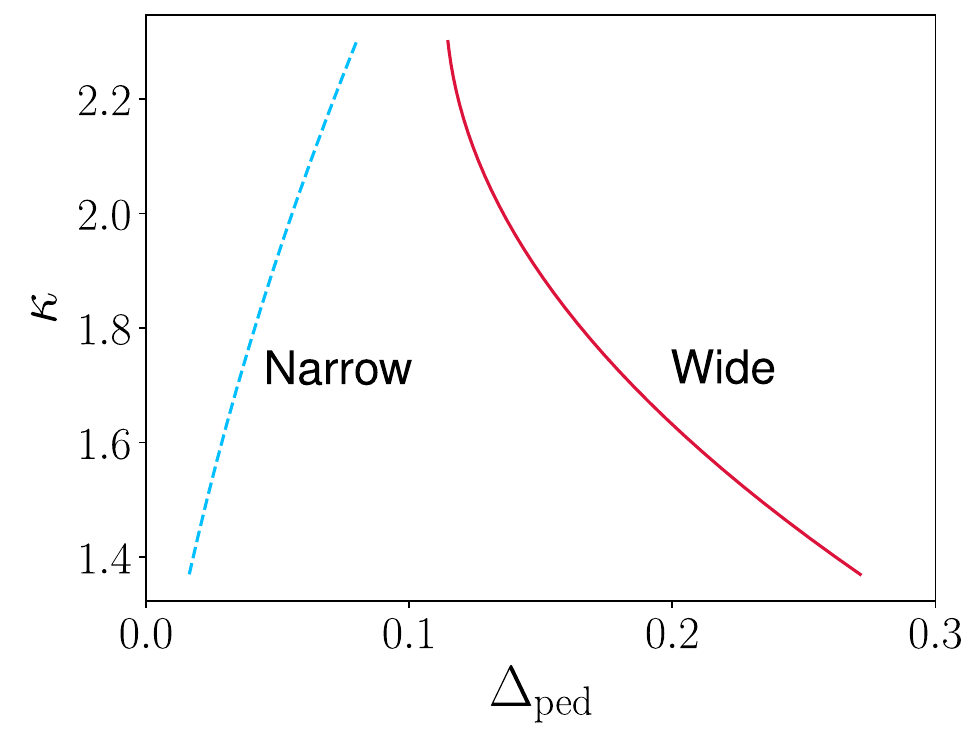}
    \caption{\textcolor{black}{Elongation bifurcation}}
    \end{subfigure}
     ~
    \begin{subfigure}[t]{0.231\textwidth}
    \centering
    \includegraphics[width=1.09\textwidth]{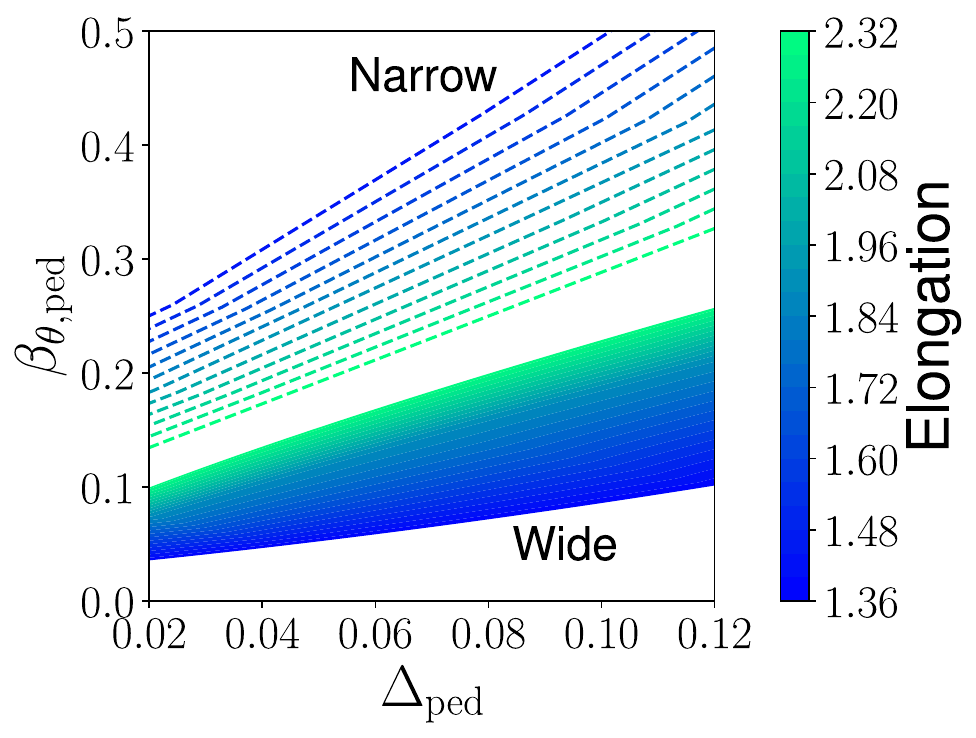}
    \caption{\textcolor{black}{Elongation scaling}}
    \end{subfigure}
    ~
    \begin{subfigure}[t]{0.23\textwidth}
    \centering
    \includegraphics[width=1.09\textwidth]{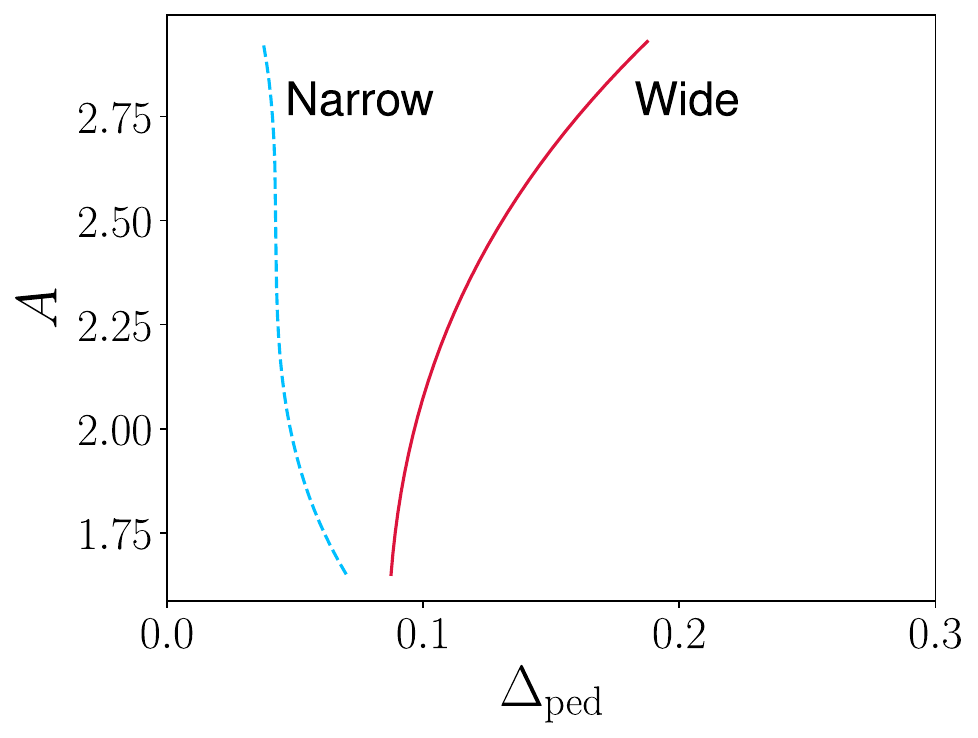}
    \caption{\textcolor{black}{Aspect-ratio bifurcation} }
    \end{subfigure}
    ~
   \begin{subfigure}[t]{0.23\textwidth}
    \centering
    \includegraphics[width=1.09\textwidth]{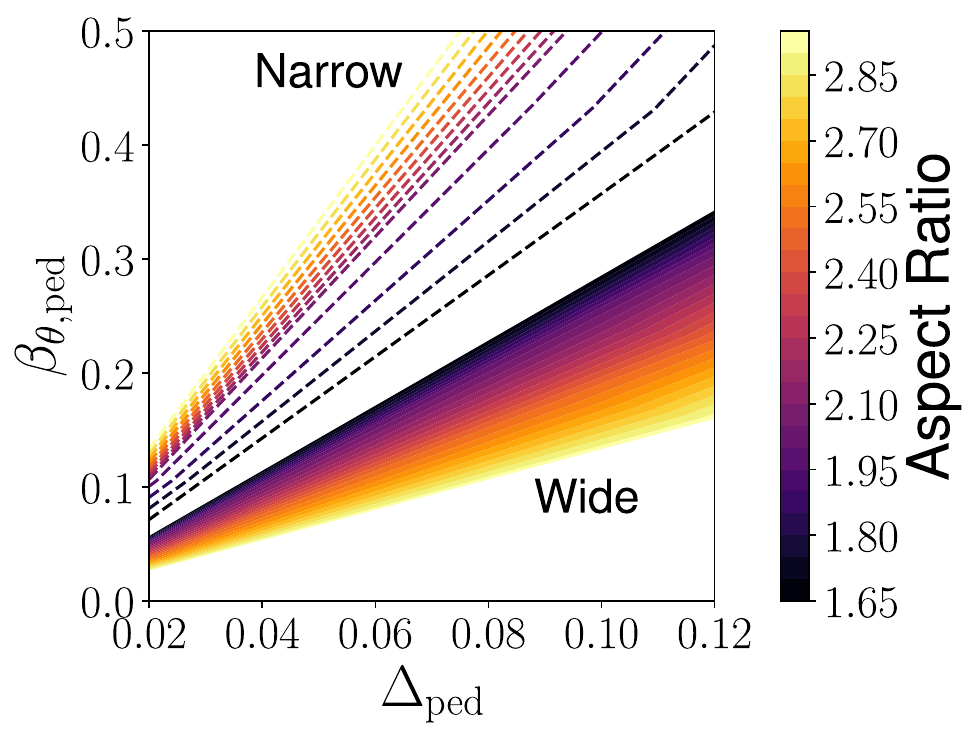}
    \caption{\textcolor{black}{Aspect-ratio scaling}}
    \end{subfigure}
    \caption{Left column: {fits (\Cref{eq:deltascalings}) for wide and narrow KBM branches} at $\beta_{\theta, \mathrm{ped}} = 0.25$ for elongation and aspect-ratio. Right column: corresponding width-height scaling expressions.}
    \label{fig:bif}
\end{figure}

In \cref{fig:four}(a) and (b), the above predictions are borne out with width-height scaling results using KBM and IBM {stability} for a triangularity scan. We emphasize that we do self-consistent equilibrium variation (outlined in \cite{Parisi2024stability}){, which changes ballooning stability as the current profile modifies $s$ and $\alpha$.} As predicted from $s-\alpha$ analysis in \cref{fig:salpha_tri_elong_aspect}, the wide branch in \cref{fig:four}(a) widens {as triangularity becomes more negative}, whereas for the narrow branch, larger positive triangularity increases the width. The wide and narrow branches scalings for the KBM are
\begin{equation}
\Delta_{\mathrm{ped, wide}} \sim 0.5^{\delta}, \;  \Delta_{\mathrm{ped, narrow}} \sim 1.7^{\delta},
\label{eq:deltascalings_tri}
\end{equation}
{plotted in \Cref{fig:four}(c).} Shown in \cref{fig:four}(b) for the IBM, while the trends for $\Delta_{\mathrm{ped}}$ are similar as the KBM in \cref{fig:four}(a), for $\delta \gtrsim -0.1$, the wide branch {disappears} for the IBM, with only the narrow branch remaining. Above {a critical} triangularity, in this example $\delta \gtrsim -0.1$, pedestal equilibria no longer access first-stability for the IBM. {This} difference in the width-height scaling between ideal and kinetic ballooning modes is important because previous treatment of the width-height scaling uses the IBM \cite{Snyder2009,Saarelma2017}, and hence, bifurcations may not always be predicted with ideal MHD.

We now show the effect of elongation and aspect-ratio on width-height scalings. In \cref{fig:bif}, the left column shows the width at constant $\beta_{\theta, \mathrm{ped}} = 0.25$, with the elongation and aspect-ratio dependence of the width scaling for the wide and narrow branches,
\begin{equation}
\begin{aligned}
& \Delta_{\mathrm{ped, wide}} \sim A^{1.5}, \;\;  \Delta_{\mathrm{ped, wide}} \sim \kappa^{-1.8} \\
& \Delta_{\mathrm{ped, narrow}} \sim A^{-0.9}, \;\; \Delta_{\mathrm{ped, narrow}} \sim \kappa^{2.9}.
\end{aligned}
\label{eq:deltascalings}
\end{equation}
In the right column of \cref{fig:bif}, we show scalings for a range of $\beta_{\theta,\mathrm{ped}}$ values. Curiously, at conventional-aspect-ratio $\gtrsim 2.5$, the wide branch has {gentle gradients close} to L-mode-like values ($-a \nabla \ln(T_e) \ll 10$), which {may} explain why no conventional-aspect-ratio H-mode experiments we have studied were in the wide-branch, despite finding wide-branch solutions in such equilibria. Notably, elongated, low-aspect-ratio plasmas, such as in NSTX, have a smaller gap in $\Delta_{\theta,\mathrm{ped}}$ between the narrow and wide branches, which might explain why {they are both} are accessible for NSTX, shown in \cref{fig:zero}.

\begin{figure}[!tb]
    \centering
    \begin{subfigure}[t]{0.5\textwidth}
    \centering
    \includegraphics[width=1.0\textwidth]{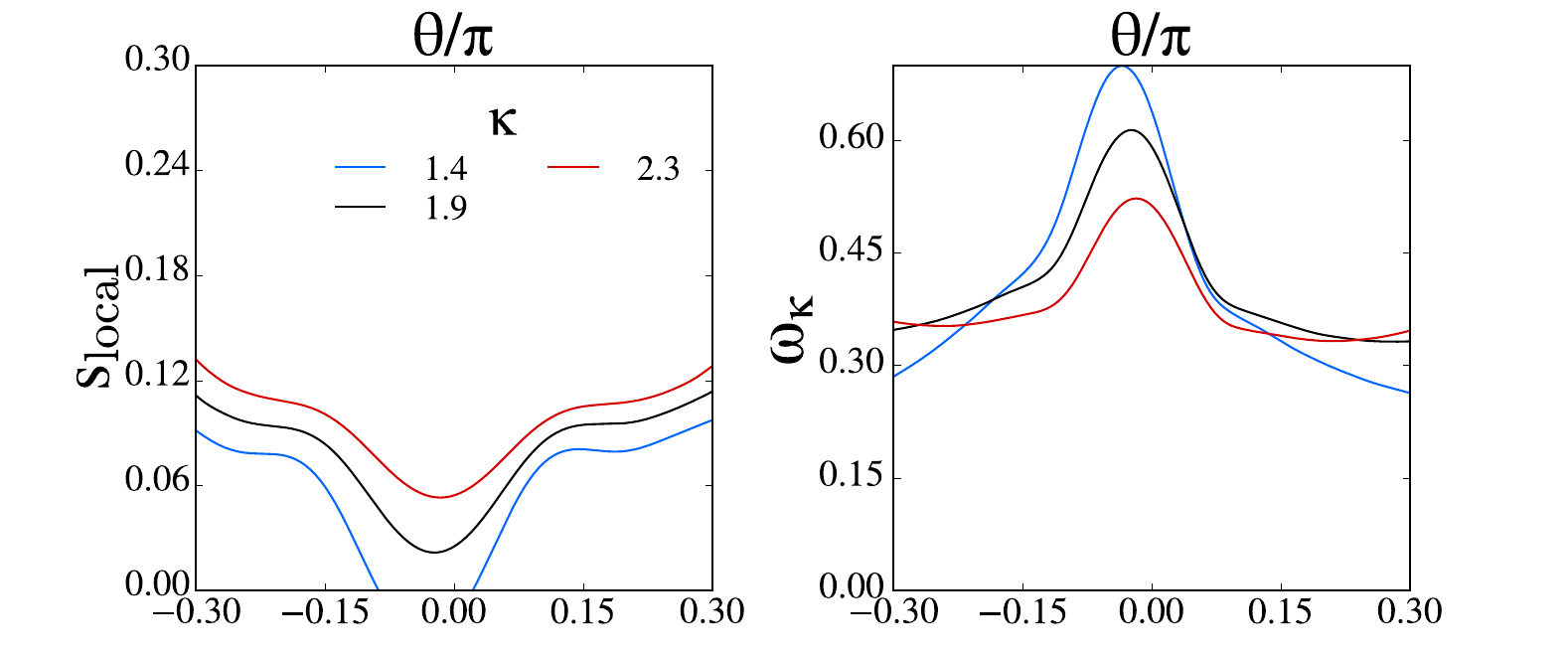}
    \caption{Elongation}
    \end{subfigure}
     ~
    \centering
    \begin{subfigure}[t]{0.5\textwidth}
    \centering
    \includegraphics[width=1.0\textwidth]{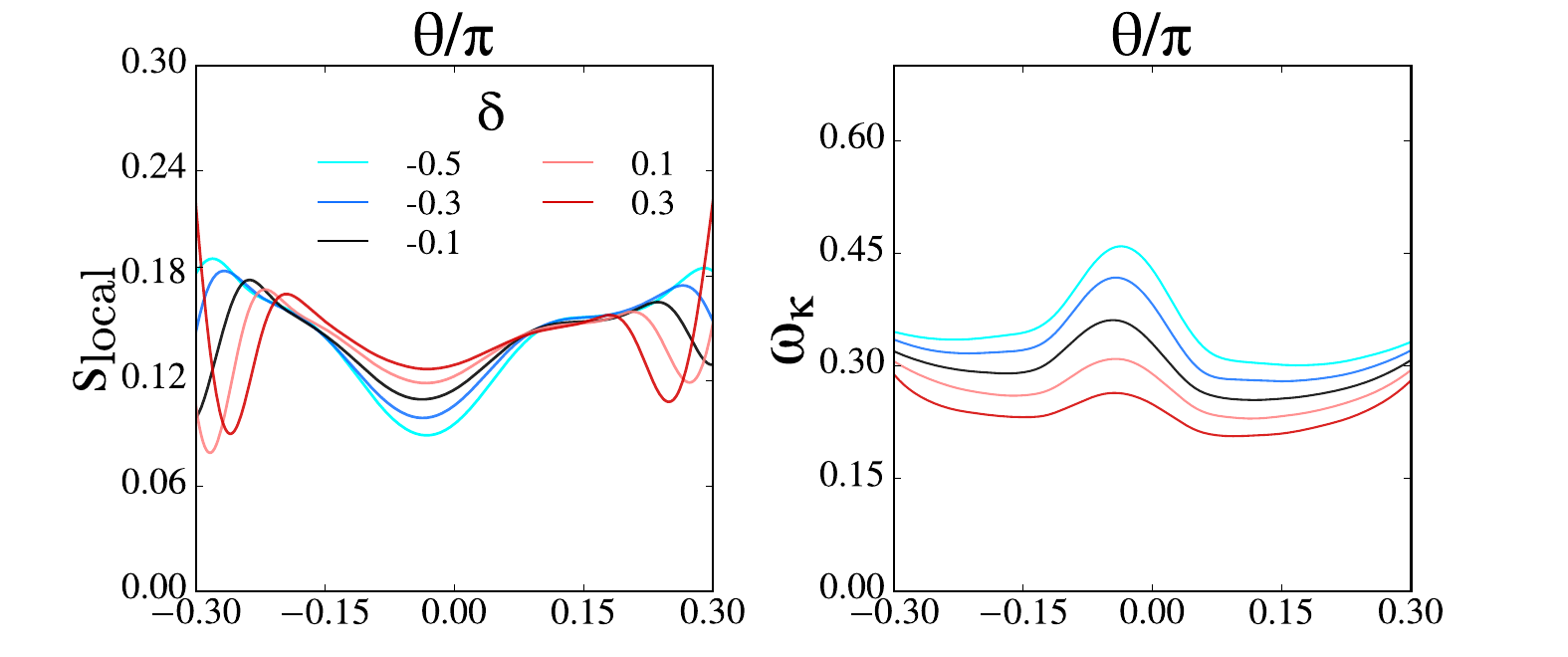}
    \caption{Triangularity}
    \end{subfigure}
     ~
    \begin{subfigure}[t]{0.5\textwidth}
    \centering
    \includegraphics[width=1.0\textwidth]{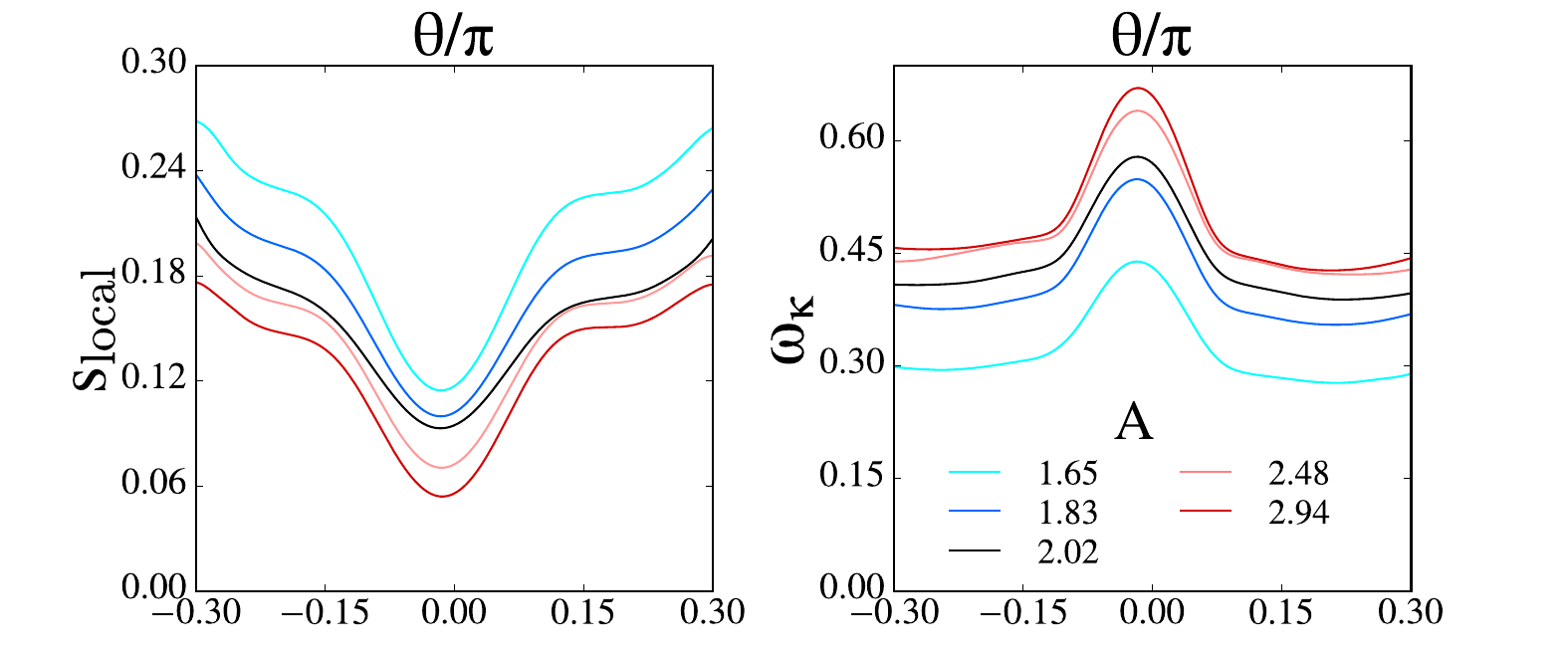}
    \caption{Aspect-Ratio}
    \end{subfigure}
    \caption{Local magnetic shear (left column) and magnetic curvature drift frequency (right column) versus poloidal angle on the mid-pedestal flux surface for (a) elongation, (b) triangularity, and (c) aspect-ratio scans.}
    \label{fig:geometry_shaping}
\end{figure}

The strong dependence of the width-height scalings on shaping and aspect-ratio can be understood by the effect of these parameters on the magnetic geometry. In \cref{fig:geometry_shaping}, we plot the local magnetic shear \cite{Greene1981},
\begin{equation}
s_{\mathrm{local}} = - \mathbf{B} \cdot \nabla \left( \frac{\nabla \Lambda \cdot \nabla \psi}{|\nabla \psi|^2} \right),
\label{eq:slocal}
\end{equation}
for the mid-pedestal flux surface, which measures the change of the safety factor in the parallel direction along a flux surface. Here, $\Lambda$ is a magnetic field-line label for a magnetic field $\mathbf{B} = \nabla \psi \times \nabla \Lambda$ \cite{Stern1970}. In \cref{fig:geometry_shaping}, we also plot the in-flux-surface magnetic curvature drift frequency,
\begin{equation}
\omega_{\kappa} = G \frac{\partial \psi}{\partial r} \left[ \left( \frac{ \mathbf{ B}}{B}  \times \nabla (B + \beta) \right) \right] \cdot \nabla \Lambda,
\label{eq:omegakappa}
\end{equation}
for {equilibria with a range of $\kappa$, $\delta$, and $A$ values but the same $\Delta_{\mathrm{ped}}$, $\beta_{\theta,\mathrm{ped}}$ values.} The {dimensionless} frequency $\omega_{\kappa}$ is normalized to the ion parallel streaming frequency using the quantity $G$. For ballooning modes, weaker magnetic shear and faster magnetic drifts tend to stabilize the modes. Crucially, local shear stabilization is sign-independent \cite{Greene1981}. Because ballooning eigenmodes peak in magnitude close to the low-field-side with poloidal angle $\theta \approx 0$, the values of $s_{\mathrm{local}}$ and $\omega_{\kappa}$ around $\theta \approx 0$ most strongly determine ballooning mode stability. \Cref{fig:geometry_shaping} shows that lower values of elongation, more negative triangularity, and higher aspect-ratio decrease the local shear and increase $\omega_{\kappa}$, all of which destabilize the KBM. This is consistent with \cref{fig:salpha_tri_elong_aspect} where, for example, in \cref{fig:salpha_tri_elong_aspect}(a), relatively small values of $\alpha$ are needed to destabilize the ballooning mode at lower elongation, which has relatively small $s_{\mathrm{local}}$ around $\theta = 0$ and very fast magnetic drifts. The {varying} impact of kinetic effects such as drift resonances and Landau damping across shaping and aspect-ratio, while included in our gyrokinetic simulations, has not been {examined closely} here, {but is an important question for future work.}

We have discovered that the width-height scaling (a) has a bifurcation, and (b) has a strong dependence on shaping and aspect-ratio. We now employ these findings to predict favorable ELM-free regimes.

\begin{figure}[!tb]
    \centering
    \includegraphics[width=0.4 \textwidth]{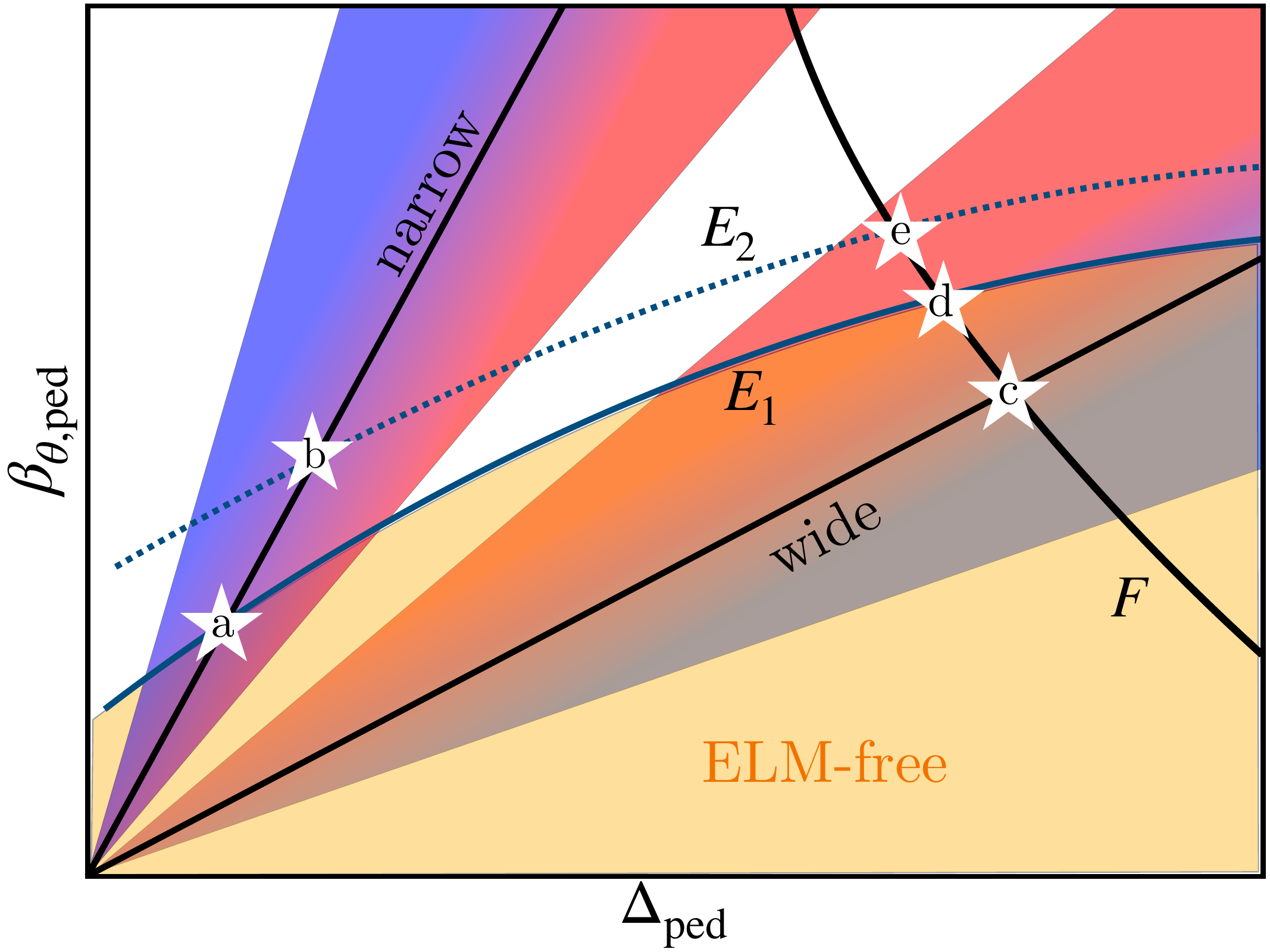}
    \caption{Schematic dependence of pedestal equilibrium points (stars) on wide and narrow KBM branches, ELM stability ($E_1$, $E_2$), and a saturation mechanism such as flow shear {degradation} ($F$). The color contours represent different shaping and aspect-ratio configurations.}
    \label{fig:three}
\end{figure}

Degraded pedestal gradients might permit ELM-free operation while achieving a high pedestal pressure \cite{Osborne2015}, which could be achieved in the wide KBM branch. In \cref{fig:three}, we schematically show the wide and narrow KBM branches, with color contours heuristically representing different shaping and aspect-ratio configurations based on results in \cref{fig:four,fig:bif}. Some possible equilibrium points are shown in \cref{fig:three}; Point (a): a conventional-aspect-ratio ELMy H-mode pedestal typically saturates where first ELM stability, curve $E_1$, intersects the narrow branch at (a). We have assumed that the ELM constraint has the scaling dependence $\Delta_{\mathrm{ped}} \sim \beta_{\theta, \mathrm{ped}}^{4/3}$ \cite{Snyder2011}. Point (b): Super H-modes \cite{Snyder2019} are obtained by {shifting the ELM stability boundary to $E_2$, resulting in a much higher pedestal at (b), but possibly still ELMing.}

It may be possible to move to a much higher $\beta_{\theta, \mathrm{ped}}$ value \textit{and} avoid ELMs by accessing the wide pedestal branch. In \cref{fig:three} we plot a new constraint $F$ {resulting from} additional transport, flow shear {degradation}, or other saturation mechanisms. $F(\Delta_{\mathrm{ped}}, \beta_{\theta, \mathrm{ped}})$ will {vary by} saturation mechanism, but preliminary analysis shows $\Delta_{\mathrm{ped}} \sim \beta_{\theta, \mathrm{ped}}^{-1}$ for a flow shear {degradation} constraint \cite{Parisi2024stability}. The intersection of the wide branch and $F$ could {give} a much higher and wider ELM-free pedestal at point (c). However, ELMy solutions may still exist on the wide branch: if an even higher $\beta_{\theta, \mathrm{ped}}$ is desired, shaping and aspect-ratio could be adjusted to move to points (d) or (e). For this illustration, we have omitted how plasma shaping and aspect-ratio changes ELM stability \cite{Snyder2015,Merle2017}. Determining the {dependence} of \textit{both} the ELM and KBM scalings on shaping and aspect-ratio, {and examining the effects of shaping and aspect-ratio with other ELM control techniques \cite{Snyder2012} is an important problem.}

{To test whether narrow and wide KBM branches are likely to ELM, we also perform ideal PBM simulations for NSTX 139047 using ELITE \cite{Snyder2002, Wilson2002, Snyder2007}. We find that the plasma is PBM-stable along wide and narrow KBM branches for mode numbers $n = 3-25$, approaching instability for very large $\Delta_{\mathrm{ped}}$ or $ \beta_{\theta, \mathrm{ped}}$ values. An important caveat is that non-ideal effects can modify PBM stability in NSTX \cite{Kleiner2021}.}

At low-aspect-ratio, negative triangularity (NT) might be particularly attractive for ELM-free H-mode operation. NT \cite{Austin2019} {has} received attention for its ELM-free and enhanced confinement characteristics \cite{Nelson2023,PazSoldan2023}. By design,  NT {does} not access H-mode, staying in first ballooning stability and L-mode \cite{Nelson2022} {to avoid ELMs}. While conventional-aspect-ratio NT {degrades} first-stable pedestals so much that edge gradients are L-mode-like \cite{Nelson2023}, at \textit{low-aspect-ratio} NT may achieve steeper gradients: \cref{fig:bif}(c) shows the wide branch has roughly twice as steep gradients at $A = 1.7$ compared with $A = 2.8$ due to $\Delta_{\mathrm{ped, wide}} \sim A^{1.5}$ (see \cref{eq:deltascalings}). Thus, a low-aspect-ratio NT H-mode could inherit the favorable properties of NT (no narrow branch access) {with H-mode-like gradients.}

{While we lack a prescription for accessing separate branches of the width-height scaling bifurcation, there is a curious observation}: a sufficiently big and sudden loss (or gain) of pedestal particles and heat -- such as an ELM (or pellet injection) -- could cause a pedestal to jump from the narrow to wide branch (or wide to narrow branch) in \cref{fig:three}. There are examples of H-modes that become ELM-free after an initial ELM \cite{Maingi2010}; jumping from narrow to wide branches is a possible explanation. Finally, companion NSTX discharges 129015 and 129038 \cite{Maingi2012}, shown in \cref{fig:zero}, fall in the the narrow and wide branches. These discharges differ mainly by Lithium dosing \cite{Kugel2008}, which {may} be responsible for different branch access. {Two other largely ELM-free regimes, I-mode \cite{Whyte2010} and wide pedestal QH-mode \cite{Chen2017}, may also access the wide pedestal branch.}

We have discovered a novel and intriguing bifurcation in the width-height scalings of tokamak pedestals, which can be modified by plasma shaping and aspect-ratio \cite{Goldston1984}. This bifurcation arises from the first and second stability properties of kinetic-ballooning-modes that give {distinct} wide and narrow pedestal branches. This discovery opens up the operating space for accessible pedestal widths and heights, offering new prospects for pedestal {regimes} in fusion experiments. While our investigations thus far have only revealed wide branch access in NSTX pedestals, investigation of how NSTX pedestals access the wide branch could provide a pathway to ELM-free operation \cite{Nelson2023} with high pedestal pressures.

We are grateful to S. C. Cowley, W. Dorland, R. Maingi, O. Sauter, P. B. Snyder, and H. R. Wilson for insightful discussions and to T. Bechtel and J. McClenaghan for technical assistance with the EFUND code. This work was supported by the U.S. Department of Energy under contract numbers DE-AC02-09CH11466, DE-SC0022270, DE-SC0022272, {DE-SC0014264, DE-SC0021629,} and the Department of Energy Early Career Research Program. The United States Government retains a non-exclusive, paid-up, irrevocable, world-wide license to publish or reproduce the published form of this manuscript, or allow others to do so, for United States Government purposes.

The data that support the findings of this study are openly available in Princeton Data Commons at \url{https://doi.org/10.34770/fc813051}. {Part of the data analysis was performed using the OMFIT integrated modeling framework \cite{OMFIT2015} using the Github project \texttt{gk\_ped} \cite{Parisi2024stability}.}

\bibliographystyle{apsrev4-1} %
\bibliography{EverythingPlasmaBib} %

\end{document}